\documentclass[10pt,conference]{IEEEtran}
\IEEEoverridecommandlockouts

\usepackage{amsmath,amssymb,amsfonts}
\usepackage{algorithmic}
\usepackage{graphicx}
\usepackage{textcomp}
\usepackage{xcolor}
\usepackage{hyperref}
\usepackage{multirow}
\usepackage{subfigure}
\usepackage[flushleft]{threeparttable}
\usepackage[authormarkup=none]{changes}
\usepackage{cite}
\usepackage{tcolorbox}

\definechangesauthor[color=red]{qing}

\graphicspath{{figures/}}
\newcommand{\figref}[1]{Fig.~\ref{#1}}
\newcommand{\tabref}[1]{Table~\ref{#1}}
\newcommand{\secref}[1]{Section~\ref{#1}}

\def\BibTeX{{\rm B\kern-.05em{\sc i\kern-.025em b}\kern-.08em
    T\kern-.1667em\lower.7ex\hbox{E}\kern-.125emX}}
    
\renewcommand\IEEEkeywordsname{Keywords}    
    
\begin{document}

\title{DeepRelease: Language-agnostic Release Notes Generation from Pull Requests of Open-source Software \\
\thanks{$^{\ast}$Corresponding author.}
}

\author{
  \IEEEauthorblockN{
      Huaxi Jiang\IEEEauthorrefmark{2}\IEEEauthorrefmark{3}, 
      Jie Zhu\IEEEauthorrefmark{2}\IEEEauthorrefmark{3},
      Li Yang\IEEEauthorrefmark{1}\IEEEauthorrefmark{2},
      Geng Liang\IEEEauthorrefmark{2},
      Chun Zuo\IEEEauthorrefmark{4}
    }
  \IEEEauthorblockA{
    \IEEEauthorrefmark{2}Institute of Software, Chinese Academy of Sciences, Beijing, China \\
    \IEEEauthorrefmark{3}University of Chinese Academy of Sciences, Beijing, China \\
    \IEEEauthorrefmark{4}Sinosoft Co.,Ltd., Beijing, China \\
    \{jianghuaxi19, zhujie212\}@mails.ucas.edu.cn\\ \{yangli2017, lianggeng\}@iscas.ac.cn \\
    zuochun@sinosoft.com.cn
  }
}

\maketitle

\begin{abstract}
The release note is an essential software artifact of open-source
software that documents crucial information about changes, such as new features and bug fixes. With the help of release notes, both developers and users could have a general understanding of the latest version without browsing the source code. However, it is a daunting
and time-consuming job for developers to produce release notes. Although
prior studies have provided some automatic approaches, they generate
release notes mainly by extracting information from code changes. 
This will result in language-specific and not being general enough to be
applicable. Therefore, helping developers produce release notes
effectively remains an unsolved challenge. To address the problem, we
first conduct a manual study on the release notes of 900 GitHub projects, which reveals that more than 54\%
of projects produce their release notes with pull requests. Based on
the empirical finding, we propose a deep learning based approach
named DeepRelease (\underline{Deep} learning based \underline{Release}
notes generator) to generate release notes according to pull requests. The process of release notes generation in DeepRelease
includes the change entries generation and the change category (i.e.,
new features or bug fixes) generation, which are formulated as a text
summarization task and a multi-class classification problem, respectively. 
Since DeepRelease fully employs text information from pull requests to
summarize changes and identify the change category, 
it is language-agnostic and can be used for projects in any language. 
We build a dataset with over 46K release notes and evaluate DeepRelease on the dataset. 
The experimental results indicate that DeepRelease outperforms four baselines 
and can generate release notes similar to those manually written ones in a fraction of the time.
\end{abstract}


\section{Introduction}
Release notes are essential software trails \cite{abebe2016empirical}, 
summarizing the primary changes that occurred since the previous release, such as new
features, bug fixes, etc. \cite{moreno_arena_2017}. Maintainers of open-source software use release notes to 
communicate with users about new changes and updates in the software system. \tabref{table:rn} presents a release note snippet from Electron\footnote{\url{https://github.com/electron/electron/releases/tag/v11.4.0}}, in which three change categories that represent high-level types of information \cite{moreno_arena_2017}, including features, fixes, and other changes. Each of the change category has several 
entries that describe the corresponding changes. In addition, at the end of each entry, there is a number that 
refers to a GitHub Pull Request (from hereon, PR), which is a distributed development model implemented by GitHub \cite{Gousios_Pinzger_Deursen_2014}. Contributors can submit their changes (e.g., implement a feature) to 
the central repository through a PR. A PR consists of a title, 
an optional free-form text that describes what changes are made and why they are needed, 
and one or more interrelated commits \cite{liu_automatic_2019}.

\begin{table}[h]
  \caption{A release note snippet from Electron.}
  \centering
\noindent \fbox{
\begin{minipage}[c][11em][c]{0.46\textwidth}
\textbf{Features}
\begin{itemize}
    \item Added support for the des-ede3 cipher in node crypto. \href{https://github.com/electron/electron/pull/27993}{\#27993}
\end{itemize}
\textbf{Fixes}
\begin{itemize}
    \item Fixed navigator.bluetooth.requestDevice crash. \href{https://github.com/electron/electron/pull/27941}{\#27941}
    \item ...
\end{itemize}
\textbf{Other Changes}
\begin{itemize}
    \item Backported fix for CVE-2020-27844. \href{https://github.com/electron/electron/pull/28101}{\#28101}
    \item ...
\end{itemize}
\end{minipage}
} 
\label{table:rn}
\end{table}


However, considering that maintainers may not understand all the changes, 
manually generating release notes can be complicated and time-consuming. 
Take open-source projects on GitHub as an example, the
general process is that one of the maintainers collects all relevant
information to create a formal release note based on the commits
\cite{bi2020tse}. Usually, each commit corresponds to a specific PR in the pull-based development model
\cite{barr2012cohesive}. Different maintainers may not have the same
level of familiarity with the project and may need to read the PRs
carefully to summarize changes, thus spending much time. Existing
study \cite{moreno_arena_2017} shows that the whole process can take
as much as eight hours. 

Some studies proposed automated approaches to generate the release
note. However, one main limitation in prior
research is that such approaches ignore text information in repositories and 
generate release notes mainly by extracting
information from code changes, which results in language-specific and
not being general enough to be applicable \cite{moreno_arena_2017,
ali2020automatic}. Furthermore, Bi et al. \cite{bi2020tse} found that
no participants in their empirical study have adopted such tools, demonstrating a clear need to develop practical tools to document release notes automatically.
Thus, how to generate release notes effectively remains an unsolved challenge.


In this paper, we propose a language-agnostic approach for open-source software to
generate release notes automatically. We first conduct a manual study 
to reveal the characteristics of release notes. By analyzing the
release notes of 900 open-source projects, we find that more than 54\%
of projects generate release notes based on PRs, and 41\% do not have the change category (i.e., new features or bug fixes). Driven by our manual study results, we propose a deep learning based
approach called DeepRelease to automatically generate release notes based on the text information in PRs. 
To the best of our knowledge, this is the first empirical step toward generating release notes from PRs using deep learning. 
The process of release notes generation in DeepRelease includes the change entries generation and the change category generation,
which are formulated as a text summarization task and a
multi-class classification problem, respectively. We evaluate DeepRelease using ROUGE for the change entries generation, and the results show that our approach outperforms the baselines in terms of ROUGE-1, ROUGE-2, and ROUGE-L by 5.0\%, 6.2\%, and 5.2\%. As for the change category classification, DeepRelease can also outperform baselines in terms of F1 score by 22.6\% at least. Overall, the experimental results indicate that DeepRelease can generate release notes similar to those manually written ones in a fraction of the time.

In summary, this paper provides the following contributions:
\begin{itemize}
  \item  We conduct a comprehensive manual study to reveal the characteristics of release notes. 
  \item We propose DeepRelease, a deep learning based and language-agnostic approach to generate release notes automatically for open-source software. It can be used for projects in any programming language.
  \item We built a dataset with over 46K release notes from GitHub, and it is publicly available\footnote{\url{https://tinyurl.com/sdcc95dj}}.
  \item We evaluate DeepRelease on the dataset, and the evaluation results show that our approach outperforms competitive baselines.
\end{itemize}

The rest of this paper is organized as follows: \secref{sec:rw}
introduces the related works and compares those works with ours. \secref{sec:manual}
presents the manual study on release notes. \secref{sec:approach} elaborates on our release notes generation approach. We describe our dataset in \secref{sec:dataset} and present the procedures and results of the
evaluation in \secref{sec:eval}. \secref{sec:threat}
discusses the threats to validity. Finally, we conclude
the paper and point out some potential directions for future research in \secref{sec:conclusion}.

\section{Related work}
\label{sec:rw}
 Software change is one of the significant characteristics of software evolution, which has different granularities like releases and commits \cite{liu_automatic_2019}.

Several empirical studies have conducted software documentation and software history investigation, manifesting that the release note is an essential artifact. Tsay et al. \cite{tsay2011experiences} collected 1579 distinct releases from 22 different open-source projects. They created a software release history database, which can help researchers and developers study and model the release engineering process in greater depth. To better understand the types of contained information in release notes, Abebe et al. \cite{abebe2016empirical} manually analyzed 85 release notes across 15 different software systems and identified six different types of information. Bi et al. \cite{bi2020tse} performed a large-scale empirical study of 32,425 release notes in 1,000 GitHub projects to understand the characteristics of real-world release notes. They also conducted interviews and an online survey to investigate professionals' opinions on release notes in practice. They found that interviewees and participants of their study mentioned that they seldom used automatic tools. Their study demonstrates a clear need for developing novel and practical tools to document release notes automatically.

Although helping developers improve release note production is a promising research direction, little research has been done on the automatic generation of release notes. Moreno et al. \cite{moreno_arena_2017} proposed ARENA (\textbf{A}utomatic \textbf{RE}lease \textbf{N}otes gener\textbf{A}tor) to generate release notes automatically for Java projects. To begin with, ARENA identifies changes that occurred in the commits, such as structural changes to the code, upgrades of external libraries used by the project, and changes in the licenses. Then, ARENA summarizes the code changes through code summarization technology. Finally, the release note is organized into categories and presented as a hierarchical HTML document. 

Similarly, Ali et al. \cite{ali2020automatic} presented an automatic approach to generating release notes for Node.js projects. There are four main modules in the proposed approach: Changes Extractor, Summarizer, Issues Extractor, and Doc Generator. First, their system extracted changes between the previous release and the new release of the project. The second step summarizes all the information get in the previous step and arranges them in a hierarchy. Then the Issues Extractor extracted all those issues added between the previous release date and the new release date from JIRA. Lastly, Doc Generator arranged all the information and generated the release notes in Docx format.

The methods mentioned above ignore text information in software repositories and 
generate release notes mainly by extracting information from code changes, 
which results in language-specific and not being universal enough to be applicable.
This may be the reason why these tools are not widely used. Different from prior studies, our approach focuses on employing text information from pull requests instead of using information from code and the specific issue tracker. Therefore it is language-agnostic and can be used for projects with the pull-based model in any programming language. Moreover, we apply a deep learning based approach to mimic the behavior of developers when making release notes. 

Researchers have also explored the summarization of the smaller granularity software changes, such as commit messages. For example, Buse and Weimer  \cite{Buse_Weimer_2010} presented a semantically rich summarization algorithm called DELTADOC, which can describe the effects of a change on program behavior, rather than simply showing what text changed. Similarly, Linares-Vasquez et al. \cite{Linares-Vasquez_Cortes-Coy_Aponte_Poshyvanyk_2015} built ChangeScribe, a tool that can automatically generate commit messages by extracting and analyzing the differences between two versions of the source code, and also performing a commit characterization based on the stereotypes of methods modified, added and removed.

Distinct from commit messages that only correspond to one commit, a release is a collection of plenty of commits and/or PRs \cite{liu_automatic_2019}. Hence they have different focuses and information structures. Besides, the existing technique for commit messages generation is limited to code changes and specific programming languages. However, there are also non-code changes in a release, like updating the documents. Thereby, the techniques for generating commit messages and release notes are complementary rather than competing with our approach. Our work summarizes the text information within PRs to generate release notes, covering code changes and non-code changes. 

\section{Manual study}
\label{sec:manual}
To better understand what information is included in release notes and examine if there exist common characteristics that can help produce release notes, in this section, we manually inspect release notes from 900 GitHub projects.

\subsection{Studied Projects}
To begin with, we use the TIOBE Programming Community Index\footnote{\url{https://www.tiobe.com/tiobe-index/}} to pick nine popular programming languages, including C, Python, Java, C++, C\#, JavaScript, PHP, Ruby, and Go. Then, because the number of stars of a GitHub repository can be seen as a proxy of its popularity \cite{borges2016understanding}, we get the top 100 repositories with the most GitHub stars in each language, for a total of 900 projects, with the help of the GitHub Ranking\footnote{\url{https://github.com/EvanLi/Github-Ranking}}.

\subsection{Manual Study Setup}
Our goal is to inspect the release notes to study their characteristics manually. For each project's release notes, we observe their style and analyze the composition of the content to search for the potential factors taking part in the generation. Specifically, this paper's first and second authors (i.e., A1 and A2) study these 900 projects and record patterns within and between projects separately. Finally, A1 and A2 compare the observed patterns and discuss any disagreement until reaching a consensus. 

\subsection{Observed Patterns}
\begin{figure}[h!]
  \centering
  \includegraphics[width=0.4\textwidth]{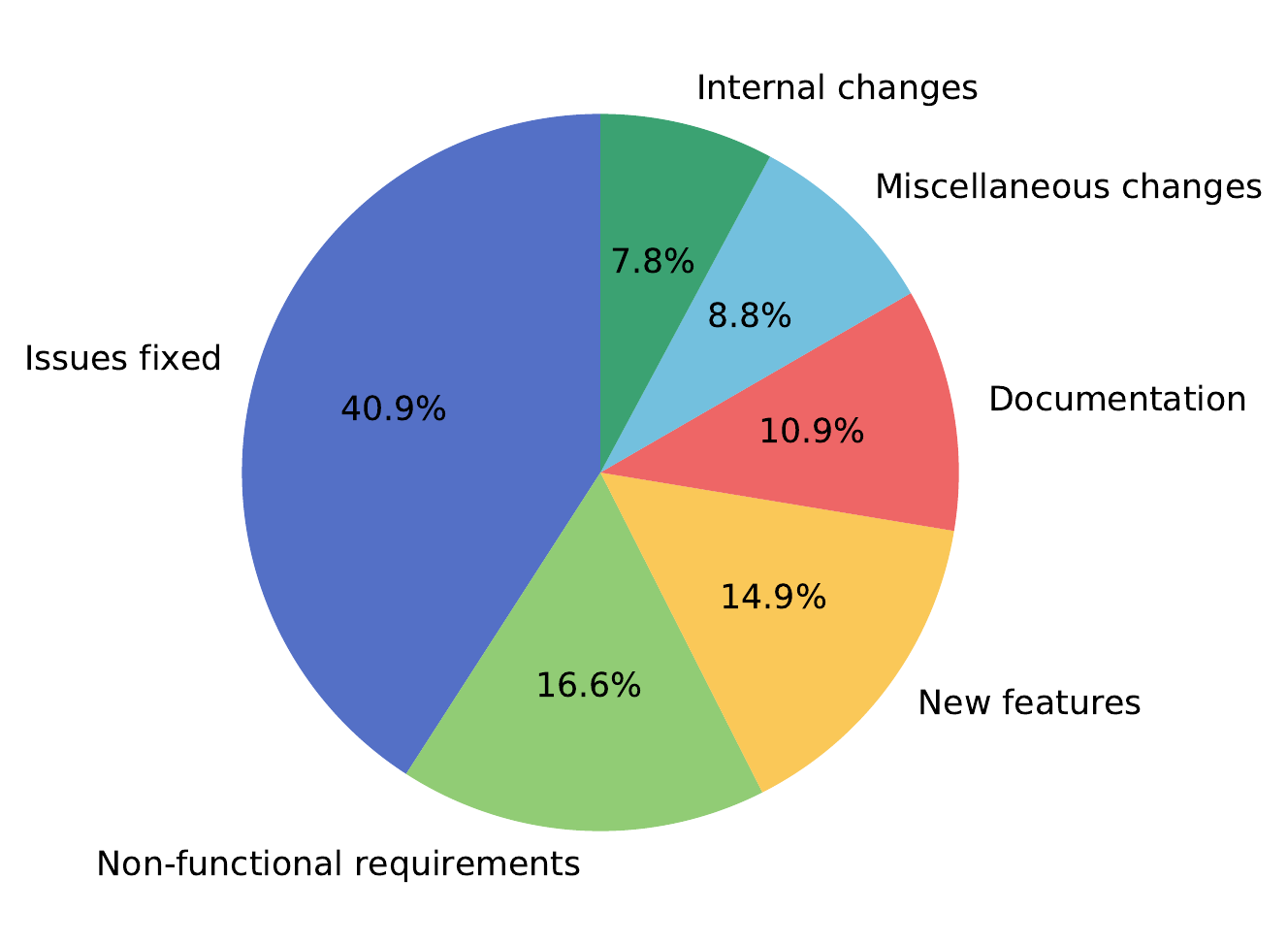}
  \caption{The percentage of categories in release notes.}
  \label{fig:category}
\end{figure}

From the horizontal aspect, we initially find that projects with and without release notes are 62.33\% and 37.67\%, respectively. Further, more than 54\% of those projects with release notes document changes with references linking to the corresponding PRs, just like the example shown in \tabref{table:rn}. Besides, we find that the release notes of different projects vary in styles and change categories due to the lack of uniform specifications. Take the change category for example, 41.9\% of projects with release notes list their software changes directly instead of organizing them by categories. For those projects with change categories, there is considerable discrepancy among their definitions of categories. For example, some projects use specific internal modules' names as change categories, and others use common categories like features and fixes, etc. Generally, all release notes we studied among these projects can be classified into six main categories based on Bi et al. \cite{bi2020tse}, \figref{fig:category} shows the percentage of each change category. 

The description of the change categories is as follows:
\begin{itemize}
    \item \textbf{Issues fixed}: Changes about what issues have been fixed.
    \item \textbf{Non-functional requirements}: Changes about a set of quality attributes of systems (for example, security and performance).
    \item \textbf{New features}: Changes about new features and functions that are newly added.
    \item \textbf{Documentation}: Changes about software artifacts, e.g., architecture documents.
    \item \textbf{Internal}: Internal changes of the systems.
    \item \textbf{Miscellaneous}: Trivial changes, e.g., update dependencies and revert changes.
\end{itemize}

\begin{figure*}[ht]
  \centering
\includegraphics[width=0.9\textwidth]{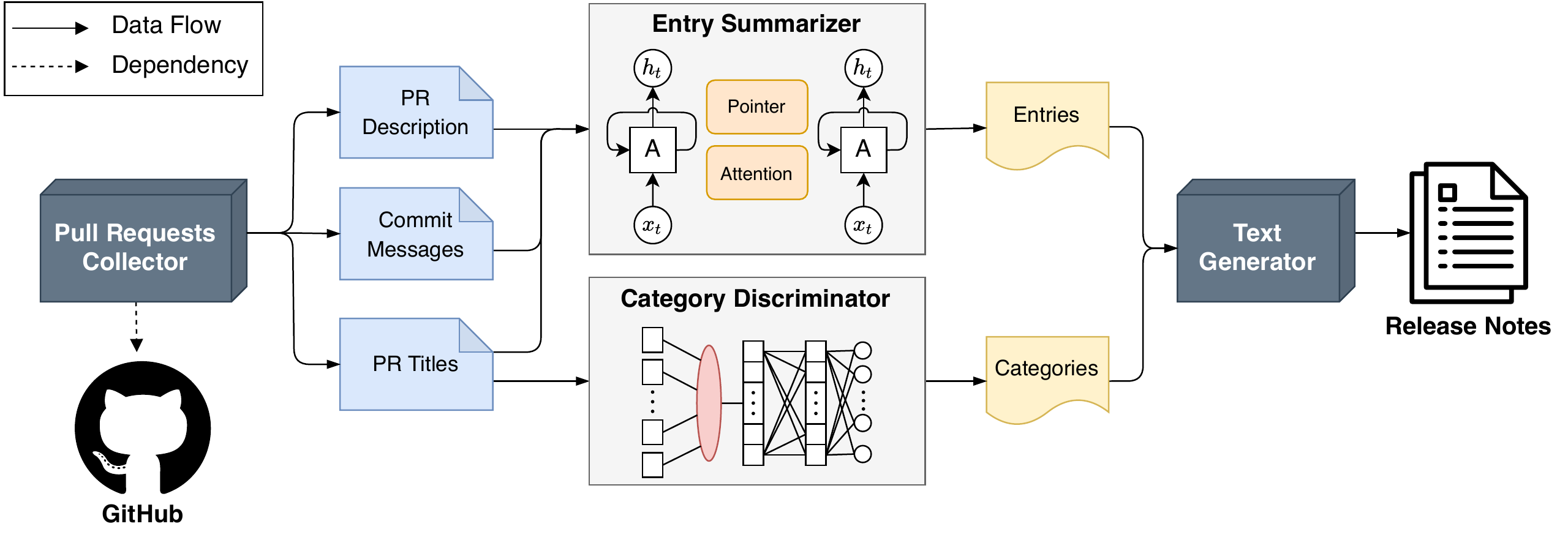}
\caption{Overview of DeepRelease.}
\label{overview}
\end{figure*}

From the vertical aspect, we find that the style of release notes can be different even within the same project. For example, some projects did not have the change category in the early release notes and later started to have them. There are two possible reasons for this. On the one hand, each release task may be undertaken by a different maintainer. On the other hand, this may be due to the fact that as the project develops and evolves, the release process gradually becomes more standardized.

\begin{center}
  \begin{tcolorbox}[colback=gray!10,
                    colframe=black,
                    width=0.48\textwidth,
                    arc=1mm, auto outer arc,
                    boxrule=0.5pt,
                   ]
In summary, our findings show that a large percentage of open-source projects do not have release notes. Furthermore, among those projects with release notes, more than 54\% of projects produce release notes based on PRs. In addition, more than 41\% of projects' release notes have no change category.
\end{tcolorbox}
\end{center}

\section{Approach}
\label{sec:approach}
As we find in \secref{sec:manual}, many projects generate release notes based on PRs and do not have change categories. Hence, PRs may contain important information that developers can employ when producing release notes. In this section, we seek to explore the potential of generating release notes with the text information of PRs. Below, we describe how we formulate problems and build deep learning models to generate release notes.

\subsection{Problem Formulation}
Based on the manual study, we denote a release note containing $n$ groups as $\boldsymbol{RN} = (G_{1}, G_{2}, ..., G_{n})$, and each group $G$ is represented as a tuple $G = <C, \boldsymbol{E}>$, where $C$ is the change category to which G belongs, and $\boldsymbol{E} = (e_{1}, e_{2}, ..., e_{m})$ denotes the change entries. Therefore, the process of generating release notes can be divided into two procedures: the generation of change entries and categories. 

First, we regard the generation of change entries as a text summarization task, which condenses a piece of text to a shorter version that contains the primary information from input document(s). The combination of the title, the description and commit messages in the PR as the source sequence $\boldsymbol{X}  = (x_{1}, x_{2}, ..., x_{|X|})$, and the entry as the target sequence $\boldsymbol{Y} = (y_{1}, y_{2}, ..., y_{|Y|})$. Therefore we need find a function $f$ so that $f(\boldsymbol{X}) = \boldsymbol{Y}$.

Second, since the number of change categories is relatively limited, we formulate the process of generating the change category as a multi-class classification problem. Given the PR title that is related to an entry, we apply a deep learning model to discriminate which category it should be.

\subsection{DeepRelease}
In this subsection, we present the overall architecture and implementation of deep learning models in DeepRelease. The overall framework of our approach DeepRelease is illustrated in \figref{overview}. It consists of four modules, i.e. Pull Requests Collector, Entry Summarizer, Category Discriminator, and Text Generator. Specifically, Pull Requests Collector firstly retrieves all URLs of PRs that merged between the previous release and the new release of the project (\secref{PRC}). For each PR, the title, the description, and commit messages are extracted and preprocessed. Then Entry Summarizer employs the PR title, the PR description, and commit messages to summarize an entry of the release note (\secref{ES}). Simultaneously, Category Discriminator uses the PR title to decide the change category, such as a new feature or a bug fix (\secref{CD}). Finally, Text Generator groups entries with the same category and generates the release notes in markdown format (\secref{DG}).

\subsubsection{Pull Requests Collector}
\label{PRC}
When a new version is about to be released, Pull Requests Collector initially retrieves all URLs of PRs that merged between the last release and the current release. For each PR, Pull Requests Collector queries its title, description, and commit messages via GitHub API. Then a series of data processing will be performed, which is described in detail in the subsection \ref{preprocess}. After that, Pull Requests Collector maintains an array of preprocessed PR, and each array element uses the number of PR to identify uniquely. These data will be sent to Entry Summarizer and Category Discriminator for further processing.

\subsubsection{Entry Summarizer}
\label{ES}
Entry Summarizer builds on sequence-to-sequence learning through a deep learning model to summarize change entries based on PRs provided by the Pull Requests Collector. Specifically, given a preprocessed PR, we concatenate its title, description, and commit messages as the input, then Entry Summarizer summarizes it to a change entry. It is pretty ubiquitous to see out-of-vocabulary (OOV) words in software artifacts due to developer-named identifiers, such as module names. To alleviate this problem, we apply the pointer generator \cite{see2017get} as Entry Summarizer, which can either generate words from the fixed vocabulary or copy words from the input sequence.

\begin{figure}[h]
    \centering
  \includegraphics[width=0.45\textwidth]{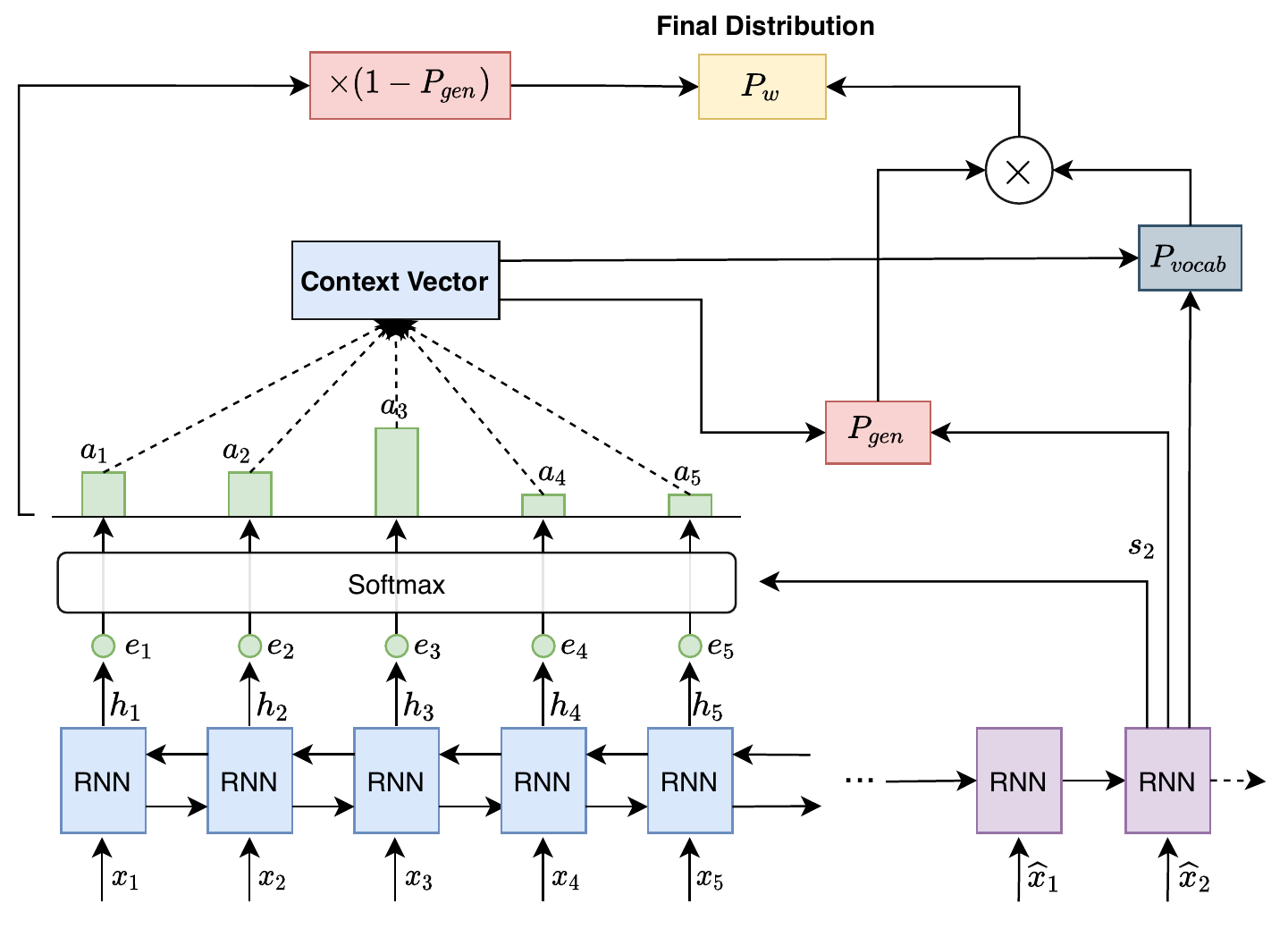}
  \caption{The pointer generator network.}
  \label{network}
\end{figure}

As shown in \figref{network}, the tokens of the source sequence $\boldsymbol{X} = (x_{1}, x_{2}, ..., x_{|X|})$ are fed one-by-one into a single-layer bidirectional LSTM\cite{hochreiter1997long} (i.e., the encoder), producing a sequence of encoder hidden states $h_{i}$. At decoding step $t$, the decoder with a single-layer unidirectional LSTM computes a decoder state $s_{t}$ and a context vector $\boldsymbol{h^{*}_{t}}$ with the attention mechanism \cite{Bahdanau15}:

\begin{align}
  e^{t}_{i} &= \boldsymbol{v}^{T}tanh(\boldsymbol{W_{h}h_{i}} + \boldsymbol{W_{s}s_{t}} + \boldsymbol{b_{e}}) \\
  \boldsymbol{a^{t}} &= softmax(e^{t}) \\
  \boldsymbol{h^{*}_{t}} &= \sum_{i}a^{t}_{i}\boldsymbol{h_{i}}  
\end{align}

\noindent where $\boldsymbol{W_{h}}$, $\boldsymbol{W_{s}}$ and $\boldsymbol{b_{e}}$ are learnable parameters. 

Then the context vector $\boldsymbol{h^{*}_{t}}$, the decoder state $s_{t}$, and the embedding $\hat{x}_{t}$ of the decoder input $\hat{y}_{t-1}$ are used to calculated the generation probability $p_{gen}$, which controls whether the word is copied from the input or generated from the vocabulary:

\begin{equation}
    p_{gen} = \sigma(\boldsymbol{W_{h^{*}}^{T}h^{*}_{t}}+ \boldsymbol{W_{s}^{T}s_{t}} + \boldsymbol{W_{x}^{T}\hat{x}_{t}} + \boldsymbol{b_{gen}})
\end{equation}

\noindent where $\boldsymbol{W_{h^{*}}^{T}}$, $\boldsymbol{W_{s}^{T}}$, $\boldsymbol{W_{x}^{T}}$, and $\boldsymbol{b_{gen}}$ are learnable parameters. $\sigma$ is the sigmoid function.

\subsubsection{Category Discriminator}
\label{CD}
\begin{figure}[h]
  \centering
\includegraphics[width=0.45\textwidth]{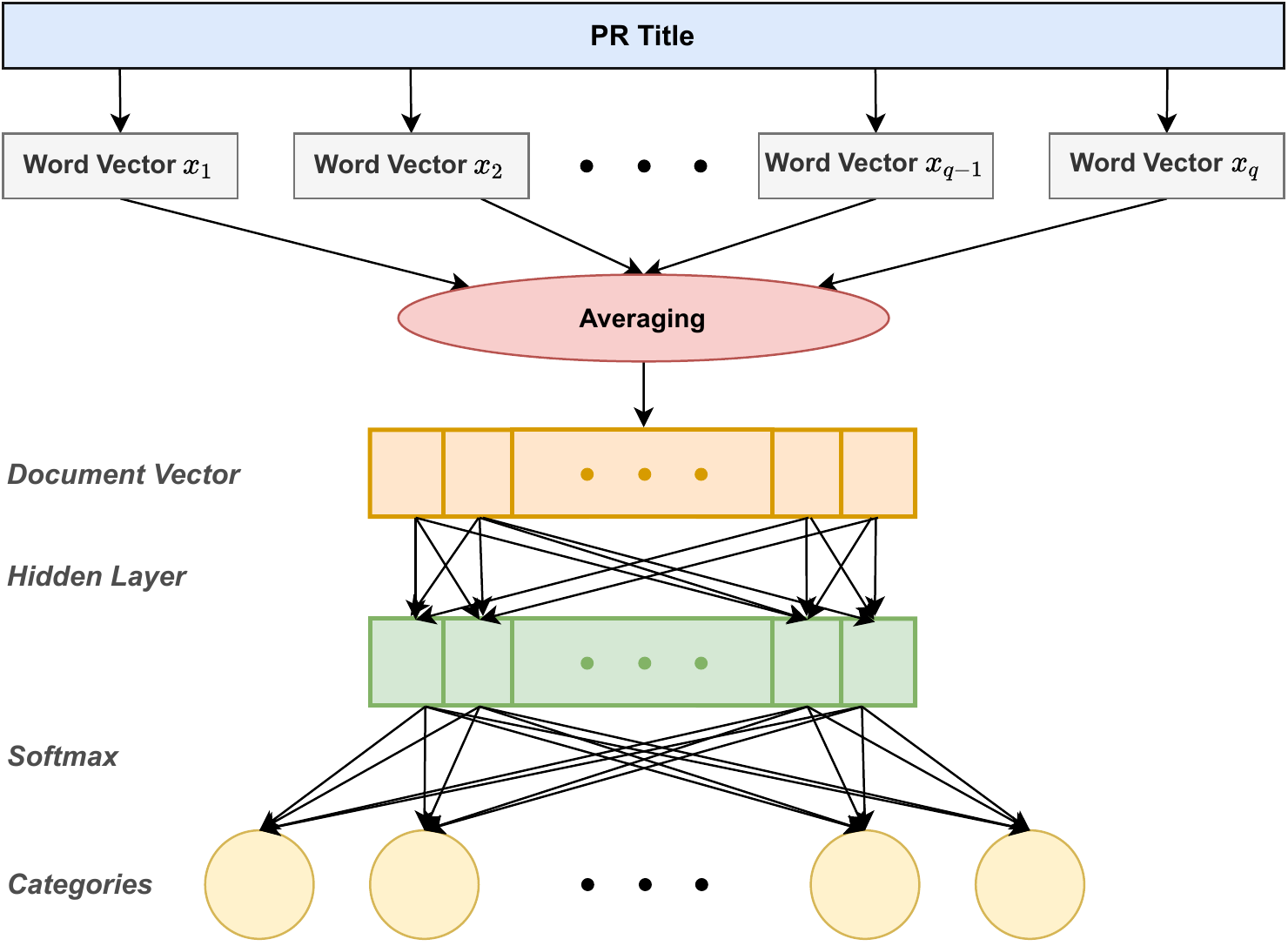}
\caption{The FastText-based multi-class classification network.}
\label{fasttext}
\end{figure}

The Category Discriminator is responsible for determining the category of entries, which can help generate release notes. For each PR passed by the Pull Requests Collector, it analyzes the PR title and decides which category the change belongs to, using a FastText-based multi-class classification model. FastText \cite{joulin2017bag} is a library for efficient learning of word representations and sentence classification. Alternative methods could be, e.g., TextCNN, TextRNN, or BERT. However, to meet the demand for faster inference and deployment, we use FastText to implement the classifier.
\figref{fasttext} depicts the structure of the multi-class classification model. Each word vector $x_{q}$ is generated by calculating the average of occurrences of the word in the PR title, and then averaged to compute the vector $y$ representing the PR title document. After that, the document vector $y$ is passed through a hidden layer where it will be multiplied by the matrix of the layer to produce a vector for classification. In the final classification step, the model uses softmax function to calculate the category probabilities $p_{k}$ by the classification vector $z$ \cite{zolotov2017analysis}:

\begin{align}
  y = \frac{1}{Q} {\textstyle \sum_{q=1}^{Q}}  x_{q} \\
  p_{k}=\frac{e^{z_{k}} }{ {\textstyle \sum_{d=1}^{D}e^{z_{d}}} } 
\end{align}

\noindent where $Q$ is the number of word vectors, $p_{k}$ is the predicted probability that the PR belongs to the $k$-th category, $z_{k}$ and $z_{d}$ are the components of the $D$ dimensional classification vector $z$.

\subsubsection{Text Generator}
\label{DG}
Text Generator is in charge of building the release note. All results of Entry Summarizer and Category Discriminator will be sent to Text Generator. It matches change entries and categories based on the number of PRs, i.e., those entries and categories with the same PR number will be paired like ``$<Category, Entry>$". Then Text Generator organizes entries by change categories and adds meta information like the version and date. Finally, the text will be rendered in markdown format. Future work on DeepRelease will analyze similar entries with the same change category and merge them via NLP technology.

\section{Dataset}
\label{sec:dataset}
\subsection{Data Collection}
\begin{figure}[h]
  \centering
  \includegraphics[width=0.45\textwidth]{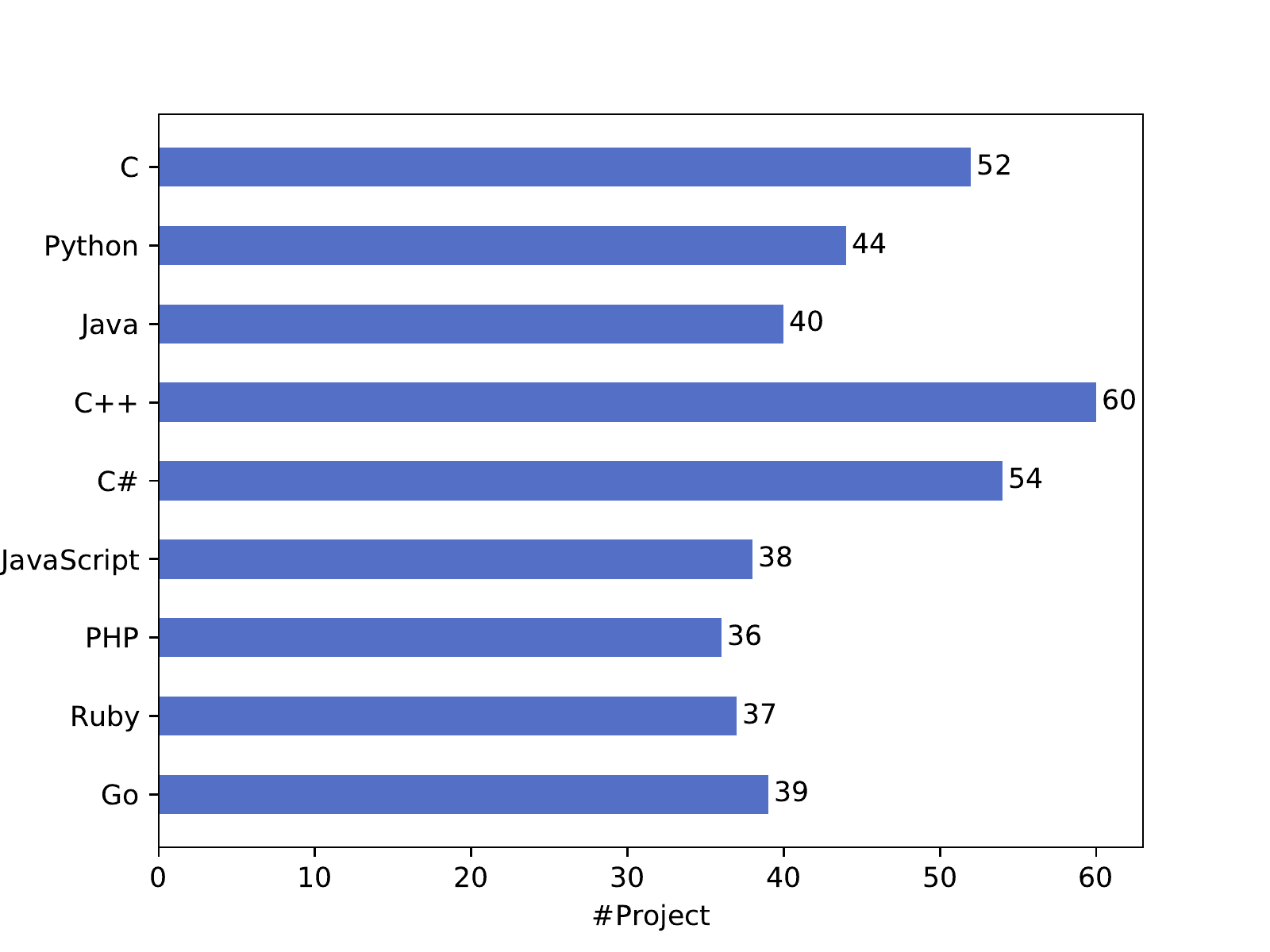}
  \caption{The number of studied projects per language.}
  \label{fig:manual1}
\end{figure}

To the best of our knowledge, there is no publicly accessible dataset of release notes with PRs and change categories. Thereby, we first select 561 projects with release notes based on the manual study to collect release notes from GitHub for DeepRelease's training. Next, to ensure that the selected projects are non-trivial, similar to a prior study \cite{bi2020tse}, we define three criteria for filtering: (1) the project is a software system rather than tutorials or books; (2) the number of release notes in the project is more than 30; and (3) the project is active and not be archived. Then we filter out 400 projects that meet our selection criteria. \figref{fig:manual1} shows that the number of studied projects per language.

For each project, we take different methods to get their release notes. Due to the lack of uniform specifications, the location of release notes varies in different projects. Some are presented on the website, some are placed as files in the project root directory, and some are published on the GitHub Releases page. Hence, we have to manually check and locate release notes and then apply different methods to collect data for different situations. For example, we use GitHub API to retrieve release notes from GitHub Releases and build crawlers for data on the web. After that, we use the regular expression to parse change categories and entries, respectively. Since DeepRelease employs the text information in PRs, for each entry, given a PR, we retrieve its corresponding PR's title, description, and commit messages via GitHub’s GraphQL API. In total, we obtain 46,656 entries of release notes with their categories and PRs from 400 repositories in different languages.

\subsection{Preprocessing}
\label{preprocess}
Our preprocessing of the collected release notes includes the following steps: 

1) Remove and replace: This procedure is to filter out trivial and templated information. For the PR description, our manual study finds that 14.8\% (133/900) of projects apply pull request templates to customize and standardize the information maintainers would like contributors to include. Hence, we query if there is a PR template in the repository for each project. If so, the template text in PR's description will be removed. Then we take the same actions to preprocess PR titles, descriptions, and commit messages. Similar to a prior study \cite{liu_automatic_2019}, sentences with URL, GitHub references (e.g., ``\#1234"), signatures, emails, mentions (e.g., ``@username"), markdown headlines, and non-ASCII characters (e.g., emoji) will be deleted. Because these tokens do not contain much information about changes, and they may bring many OOV words to interfere with our model. To further reduce the number of unique tokens, we convert SHA1 hash digest, version strings, and numbers into ``sha", ``version", and 0, respectively.

2) Unify the change category: As we find in the manual study, different projects have different definitions of change categories. For example, for those changes that fix problems, some projects used ``Fix", and others used ``Bug fixes", even though they mean the same thing. Therefore, we apply a heuristic approach to unify these categories. We begin by setting some category keywords in advance, which indicate what the entry belongs to. \tabref{tab:kws} shows the keywords we used. Take the category ``Issues fixed" as an example, if the category we extracted from release notes contains any keywords of ``fix", ``bug", and ``issue" (ignore case), it will be renamed to ``Issues fixed". Hence, categories like ``Fixes", ``Bug Fixes" will be unified to ``Issues fixed". To simplify the problem, we only consider the top-4 categories in our manual study, i.e. ``Issues fixed", ``Non-functional", ``New features", and ``Documentation". The reason is two-fold. On the one hand, they have the largest amount of data and are more balanced, making them more suitable for model learning. On the other hand, we focus on the method's versatility since this is the first work in the literature applying deep learning to classify release note entries. Thus, we exclude those categories that are closely related to specific projects. Extending our model to classify more change categories is part of our future work.

\begin{table}[h]
\centering
\caption{The keywords of each change category.}
\label{tab:kws}
\resizebox{0.45\textwidth}{!}{%
\begin{tabular}{l|l}
\hline
\textbf{Category} & \textbf{Keywords}                                                                            \\ \hline
Issues fixed      & fix, bug, issue                                                                              \\ \hline
Non-functional    & \begin{tabular}[c]{@{}l@{}}enhance, improve, performance, \\ optimize, security\end{tabular} \\ \hline
New features      & add, feature, new                                                                            \\ \hline
Documentation     & doc, documentation, wiki                                                                     \\ \hline
\end{tabular}%
}
\end{table}

3) Construct the input for models: For the task of text summarization, we concatenate the title, description, and commit messages for each PR as the input. The commit messages are sorted by their creation time in ascending order. These three sections are separated by a particular token "[sep]". For the multi-class classification problem, we just remove punctuation at the beginning and end of the PR title.


\section{Evaluation}
\label{sec:eval}
In this section, we introduce the evaluation metrics and the baselines firstly. Next, we describe our experiment settings and research questions. Lastly, we show the corresponding experimental results.

\subsection{Evaluation Metrics}
We evaluate the Entry Summarizer and its baselines with the ROUGE metric \cite{Lin_2004}, reporting the scores for ROUGE-1, ROUGE-2, and ROUGE-L that respectively measure the word-overlap, bigram-overlap, and the longest common sequence between the reference summary created by humans and the summary to be evaluated. The ROUGE metric has been shown to be highly relevant to human assessments of summarized text quality \cite{Lin_2004} and is widely used for evaluating text summarization tasks\cite{see2017get,paulus2017deep}. The precision, recall, and F1 score for ROUGE-N are calculated as follows:

\begin{align}
  P_{rouge-n} &= \frac{\sum_{X, Y\in S}\sum_{gram_{n}\in X}Cnt_{Y}(gram_{n})}{\sum_{X, Y\in S}\sum_{gram_{n}\in Y}Cnt_{Y}(gram_{n})} \\
  R_{rouge-n} &= \frac{\sum_{X, Y\in S}\sum_{gram_{n}\in X}Cnt_{Y}(gram_{n})}{\sum_{X, Y\in S}\sum_{gram_{n}\in X}Cnt_{X}(gram_{n})} \\
  F1_{rouge-n} &= \frac{2P_{rouge-n}R_{rouge-n}}{P_{rouge-n} + R_{rouge-n}}
\end{align}

\noindent where $X$, $Y$, and $S$ refer to a reference summary, a generated summary, and the test set. $gram_{n}$ is an n-gram phase, and $Cnt_{X}(gram_{n})$ and $Cnt_{Y}(gram_{n})$ refer to the occurrence number of $gram_{n}$ in $X$ and $Y$, respectively.

Similarly, ROUGE-L is computed as follows:

\begin{align}
  P_{rouge-l} &= \frac{LCS(X, Y)}{length(X)} \\
  R_{rouge-l} &= \frac{LCS(X, Y)}{length(Y)} \\
  F1_{rouge-l} &= \frac{2P_{rouge-l}R_{rouge-l}}{P_{rouge-l} + R_{rouge-l}}
\end{align}

\noindent where $LCS(X, Y)$ is the length of the longest common subsequence of $X$ and $Y$.

\begin{table*}[ht]
    \centering
    \caption{Comparisons of our approach with each baseline for the entry summarization task.}
    \label{table:res}
    \begin{tabular}{c|ccc|ccc|ccc}
    \hline
    \multirow{2}{*}{Approach} & \multicolumn{3}{c|}{ROUGE-1}                       & \multicolumn{3}{c|}{ROUGE-2}                       & \multicolumn{3}{c}{ROUGE-L}                        \\ \cline{2-10} 
                              & P              & R              & F1               & P              & R              & F1               & P              & R              & F1               \\ \hline
    LeadCM                    & 48.23          & 56.01          & 49.28            & 36.36          & 44.80          & 37.85            & 48.71          & 56.62          & 50.17            \\
    PR-Title                  & 71.26          & 71.80          & 69.73            & 60.20          & 63.45          & 60.44            & 70.49          & 72.31          & 69.88            \\
    DeepRelease               & \textbf{77.38} & \textbf{71.97} & \textbf{73.21}   & \textbf{66.54} & \textbf{63.62} & \textbf{64.21}   & \textbf{76.94} & \textbf{72.40} & \textbf{73.51}   \\ \hline
    Ours vs LeadCM            & 60.44\%        & 28.49\%        & \textbf{48.56\%} & 83.00\%        & 42.01\%        & \textbf{69.64\%} & 57.96\%        & 27.87\%        & \textbf{46.52\%} \\
    Ours vs PR-Title          & 8.59\%         & 0.24\%         & \textbf{4.99\%}  & 10.53\%        & 0.27\%         & \textbf{6.24\%}  & 9.15\%         & 0.12\%         & \textbf{5.19\%}  \\ \hline
    \end{tabular}
\end{table*}  

For the multi-class classification task, we use Precision, Recall, and F-measure as our evaluation metrics. The precision is intuitively the ability of the classifier not to label as positive a sample that is negative, and the recall is intuitively the ability of the classifier to find all the positive samples. F1 score balances the precision and recall, and provides a more realistic measure of the performance by using both of them. 



\subsection{Baselines}
For the task of generating release notes' entries, we use two baselines: LeadCM and PR-Title. LeadCM outputs the first few tokens of the commit messages in a PR as a summary. It was firstly proposed by Liu et al. \cite{liu_automatic_2019}, they hypothesized that developers might commit key changes first and make other trivial changes later. Thereby, the first few commit messages may summarize the primary information of a PR. In this paper, we make LeadCM output the first 9 tokens, because 9 is the average length of entries in our dataset. In addition, we propose PR-Title for this task. Given a PR, as the name implies, PR-Title uses the PR title as the entry. The hypothesis behind PR-Title is that developers may be inclined to use a concise sentence as a title to summarize the change in a PR.

\begin{table*}[t]
  \begin{threeparttable}
  \centering
      \caption{Comparisons of our approach with each baseline for the category classification task.}
      \label{table:clf_res}
  \begin{tabular}{c|c|ccc|ccc|ccc}
  \hline
  \multirow{2}{*}{Category} & \multirow{2}{*}{Quantity} & \multicolumn{3}{c|}{Precision}    & \multicolumn{3}{c|}{Recall}             & \multicolumn{3}{c}{F1}           \\ \cline{3-11} 
                            &                           & RG    & Keywords & DeepRelease    & RG    & Keywords       & DeepRelease    & RG    & Keywords & DeepRelease    \\ \hline
  Issues fixed              & 18564                       & 10.62 & 57.29    & \textbf{80.19} & 10.50 & 20.04          & \textbf{78.87} & 10.56 & 29.69    & \textbf{79.52} \\
  Non-functional            & 8481                      & 48.02 & 59.57    & \textbf{86.69} & 47.95 & \textbf{90.34} & 89.38          & 47.98 & 71.80    & \textbf{88.01} \\
  New features              & 7311                       & 19.34 & 45.11    & \textbf{78.81} & 19.00 & 46.08          & \textbf{86.29} & 19.17 & 45.59    & \textbf{82.38} \\
  Documentation             & 4209                      & 21.30 & 44.10    & \textbf{74.78} & 21.82 & 6.64           & \textbf{63.74} & 21.55 & 11.54    & \textbf{68.82} \\ \hline
  \end{tabular}
  \begin{tablenotes}
        \small
        \item Note: Quantity is the number of each category in the entire dataset.
      \end{tablenotes}
    \end{threeparttable}
    \end{table*}

To classify the change category, we use Random Guess (RG) and keywords-based heuristics as our baselines. Random Guess is commonly used by prior studies \cite{Li_Chen_Shang_2020, chen2014detecting, li2019dlfinder, liu2019variables,valdivia2014characterizing}. Given a PR title, it discriminates which category the entry should belong to based on the proportion of change categories in the dataset. We repeat the Random Guess 30 times (as suggested by previous studies \cite{chen2014detecting,georges2007statistically,Li_Chen_Shang_2020}) for each input to reduce the biases, and use the average values as the result. In addition, we propose a keywords-based heuristics that discriminates the change category based on keywords in the PR's title. The keywords we set are the same as in the \tabref{tab:kws}.

\subsection{Experiment Settings}
For the task of generating release notes' entries, we randomly split the 46K release notes in the ratio of 8:1:1 for training, validation, and testing, similar to the prior work \cite{jiang_ase_2017,liu_automatic_2019}. Similar to Liu et al. \cite{liu_automatic_2019}, we use 128-dimensional word embeddings. The encoder is a single-layer bidirectional LSTM and the decoder is the same but unidirectional. Both of them use 256-dimensional hidden states. The vocabulary size is set to 30K. When training the pointer generator network, we leverage Adam \cite{adam} with a batch size of 8. The beam size is set to 4 when the model generates sequences in testing.

As for the task of multi-class classification, we firstly employ some data augmentation techniques \cite{wei-zou-2019-eda} on the original dataset. Then the dataset is split in the ratio of 7:1 for training and testing. Each sentence in the training set of Category Discriminator is slightly modified using four specified operations, including synonym replacement, random swap, random insert, and random deletion. Through this method, the training set grows from 42k items to more than 420k items. Then we further train our Category Discriminator with the enhanced dataset using 100-dimensional word vectors. 

\subsection{Research Questions}
Our evaluation aims at answering the research questions described in the following paragraphs.
 
\textbf{RQ1: How effective are DeepRelease when generating the change entries and categories?} We investigate the performance of our deep learning models in this RQ. Our finding may help prove that there is sufficient text information that can be used to generate release notes in PRs. Specifically, we split the RQ into two sub-RQs:

\noindent RQ1.1: Does Entry Summarizer outperform the baselines?

\noindent RQ1.2: Does Category Discriminator outperform the baselines?

\textbf{RQ2: Is DeepRelease language-agnostic?}  We assess whether DeepRelease is a language-agnostic solution for generating release notes via qualitative analysis. 
Since ARENA \cite{moreno_arena_2017} and Ali’s method \cite{ali2020automatic} are not available, and they are evaluated manually, it is hard to do quantitative analysis with them.

\textbf{RQ3: What is the inference time of models in DeepRelease?} This RQ focus on the efficiency and applicability of our models. When generating release notes automatically, it is unreasonable to expect the developer to spend a similar amount of time that generating manually costs. Therefore, we perform a timing analysis to assess the time needed to generate entries and change categories. Precisely, we separately measure the model's inference time it takes for the number of PRs to go from 10 to 100 with an increment of 10. Then we calculate the average time required to generate each entry and change category. It is worth noting that training time is not considered since this is a one-time cost.

\subsection{Results}
\textbf{RQ1.1: Does Entry Summarizer outperform the baselines?}
\tabref{table:res} shows the evaluation results. First, we can find that our approach outperforms LeadCM and PR-Title in terms of all metrics, indicating that our approach can summarize the entry of release notes more precisely than the two baselines. Second, although the F1 scores for ROUGE are typically between 0.2 to 0.4 in text generation tasks \cite{see2017get,paulus2017deep,wan2018improving,liu_automatic_2019}, DeepRelease can reach 0.6 to 0.7, which reveals that the number of overlapping units between the generated entry and the entry created by developers is high.

We argue that the reason our approach achieves better performance is two-fold. On the one hand, our approach learns knowledge about which phrases or sentences are crucial for summarizing an entry from a large amount of historical data. On the other hand, the pointer generator model can generate novel words and copy words from the input. It can generate entries with different lengths based on the input and rephrase essential sentences. This advantage of DeepRelease reduces the number of trivial tokens and results in high precision. On the contrary, LeadCM and PR-Title cannot rephrase extracted sentences, generate novel words, or dynamically decide how many tokens to generate, since they generate entries by extracting sentences.

Third, it is interesting to note that the complicated DeepRelease outperforms the naive PR-Title by relatively small margins (from 4.99\% to 6.24\%) than the LeadCM. In particular, the recalls of our approach are very close to that of PR-Title (from 0.12\% to 0.27\%). We argue that there are two main reasons. The one is the hypothesis of PR-Title, which is that developers may tend to use a concise sentence as the title to summarize the PR. Hence, phrases produced by PR-Title may contain some tokens in the reference summary and improve the recall. The other is probably because some projects use tools that directly select the PR's title as an entry to generate release notes automatically.

\begin{center}
  \begin{tcolorbox}[colback=gray!10,
                    colframe=black,
                    width=0.48\textwidth,
                    arc=1mm, auto outer arc,
                    boxrule=0.5pt,
                   ]
In summary, our approach outperforms the two baselines in terms of ROUGE and can generate entries resembling those manually written by developers.
  \end{tcolorbox}
  \end{center}

\textbf{RQ1.2: Does Category Discriminator outperform the baselines?}
\tabref{table:clf_res} presents the evaluation results of multi-class classification. Overall, Category Discriminator outperforms the baselines by significant margins for nearly all metrics, except the recall of the Non-functional category. This may be because it is easier to dig out more representative keywords of Non-functional changes than other categories, making it have a broader match. In addition, although the recall of the Non-functional category is slightly lower, DeepRelease could achieve a much higher F1 score because of its better precision. 

We also find that the metrics on the Documentation category are much lower than other categories. There are two likely causes: 1) Insufficient amount of data limits the model's performance; 2) Developers may use the specific names of documents to introduce the change so that DeepRelease can not figure it out.
\begin{center}
  \begin{tcolorbox}[colback=gray!10,
                    colframe=black,
                    width=0.48\textwidth,
                    arc=1mm, auto outer arc,
                    boxrule=0.5pt,
                   ]
DeepRelease is able to outperform two baselines significantly in classifying four common change categories.
\end{tcolorbox}
\end{center}

\textbf{RQ2: Is DeepRelease language-agnostic?}
\tabref{tab:rq1} shows the comparison result of supported languages between DeepRelease and two related works. ARENA is currently implemented to work with Java-based systems, and it depends on a Java code analyzer to parse the source code \cite{moreno_arena_2017}. Similarly, Ali's method analyze code changes with the help of Abstract Syntax Tree (AST). Thereby, these approaches are language-specific and adapting them to work with projects implemented in other programming languages requires additional engineering effort. Whereas DeepRelease is trained on release notes from C, Python, Java, C++, C\#, JavaScript, PHP, Ruby, and Go projects in GitHub, and achieves promising results on generating release notes. We select the nine programming languages according to the TIOBE Programming Community Index, which is an indicator of the popularity of programming languages. 
  
\begin{table}[h]
    \centering
    \caption{Comparison of supported programming languages}
    \label{tab:rq1}
    \begin{tabular}{c|c|c|c}
    \hline
    \textbf{Language}   & \textbf{ARENA\cite{moreno_arena_2017}}    & \textbf{Ali's \cite{ali2020automatic}}   & \textbf{DeepRelease} \\ \hline
    C          & -        & -        & $\surd$    \\ \hline
    Python     & -        & -        & $\surd$    \\ \hline
    Java       & $\surd$ & -        & $\surd$    \\ \hline
    C++        & -        & -        & $\surd$    \\ \hline
    C\#         & -        & -        & $\surd$    \\ \hline
    JavaScript & -        & $\surd$ & $\surd$    \\ \hline
    PHP        & -        & -        & $\surd$    \\ \hline
    Ruby       & -        & -        & $\surd$    \\ \hline
    Go         & -        & -        & $\surd$   \\ \hline
    \end{tabular}
    \end{table}
  
Note that this does not mean that DeepRelease only supports these nine languages, because we use natural language information in PRs as input instead of source code. DeepRelease learns general knowledge from PRs about how to produce release notes thus achieves language-independent.
  
 \begin{center}
  \begin{tcolorbox}[colback=gray!10,
                    colframe=black,
                    width=0.48\textwidth,
                    arc=1mm, auto outer arc,
                    boxrule=0.5pt,
                   ]
DeepRelease is language-agnostic and can be used for projects in any programming language without additional engineering effort.
\end{tcolorbox}
\end{center}

\textbf{RQ3: What is the inference time of models in DeepRelease?}

\begin{table}[h]
\centering
\caption{Milliseconds per Entry/Category Generation.}
\label{tab:time}
\begin{tabular}{c|c|c}
\hline
\textbf{\#PR} & \textbf{Entry Summarizer} & \textbf{Category Discrimator} \\ \hline
10           & 26.830                    & 0.017                         \\
20           & 28.950                    & 0.016                         \\
30           & 32.150                    & 0.014                         \\
40           & 28.682                    & 0.012                         \\
50           & 29.272                    & 0.012                         \\
60           & 31.662                    & 0.012                         \\
70           & 30.323                    & 0.012                         \\
80           & 31.198                    & 0.012                         \\
90           & 30.128                    & 0.012                         \\
100          & 27.795                    & 0.012                         \\ \hline
Average      & 29.699                    & 0.013                         \\ \hline
\end{tabular}
\end{table}

Given the specified number of inputs, we compute the average inference time for each output, as shown in \tabref{tab:time}. The reported time is in milliseconds for each provided input. For instance, it would take on average about 26.83 milliseconds to generate an entry and 0.017 milliseconds to identify the change category when the input size is 10 PRs. On average, it takes 29.699 ms and 0.013 ms to generate each entry and category, respectively. These results are repeated ten times and computed using a single NVIDIA RTX 3090 GPU. It is apparent from this table that it is quick for models to infer, demonstrating the usability and effectiveness of DeepRelease.

\begin{center}
  \begin{tcolorbox}[colback=gray!10,
                    colframe=black,
                    width=0.48\textwidth,
                    arc=1mm, auto outer arc,
                    boxrule=0.5pt,
                   ]
We find that DeepRelease can generate an entry and its change category in an average of 29.712 milliseconds.
  \end{tcolorbox}
  \end{center}

\section{Threats to validity}
\label{sec:threat}
In this section, we want to acknowledge several threats to the paper's validity, such as the factors that can bias our study. These factors can be classified into three categories: construct validity, internal validity, and external validity.

Construct validity concerns the relationship between the theory and the observation. Our approach presumes that the open-source software employs the pull-based develop model and all PRs follow good practice. However, no industrial standards guide developers to write PR titles, descriptions and labels, etc. Our objective in this paper is to generate release notes that can be learned from the history of repositories. Further improvement on human-written PRs falls outside the scope of this paper.

Internal validity threats concern factors that are internal to our study that could influence our results. We conduct manual studies to investigate the characteristics and uncover the patterns of release notes. The authors examine the data independently to avoid biases. Any disagreement is discussed until a consensus is reached. Involving third-party open-source developers to verify our results might further reduce this threat. 

Different parameters used in the neural networks might affect the effectiveness of the trained models as well. We follow prior studies \cite{see2017get,liu_automatic_2019,joulin2017bag} to set parameters for our deep learning models. The models trained using our approach might not be optimal on some of the evaluation metrics. Future study will further improve the performance of DeepRelease. 

Finally, external validity is related to the generalizability of our findings. We conduct our study only on 900 open-source systems. However, we select the studied systems in various domains and programming languages to improve our studied systems' representativeness. Note that we did not compare DeepRelease with previous approaches such as ARENA, since they are not publicly available, which would require additional engineering effort to implement them. Further, our main goal is to empirically investigate the feasibility of a learning-based and language-agnostic approach for generating release notes.

\section{Conclusions and future work}
\label{sec:conclusion}
This paper aims to tackle the challenge of generating release notes for open-source projects in a more general way. We reveal some characteristics of release notes and find that more than 54\% of projects produce their release notes with pull requests by conducting a manual study. Based on the finding, we propose a deep learning based approach named DeepRelease to utilize text information in pull requests for the generation of release notes. Since DeepRelease only requires texts, it is language-agnostic and more general. The process of generating release notes in DeepRelease includes generating entries and change categories, which are formulated as a text summarization task and a multi-class classification problem. 
DeepRelease outperforms competitive baselines and achieves promising results on both summarizing the entry and discriminating the change category. The experimental results highlight the potential of generating release notes by leveraging text information in pull requests. We present that a learning-based approach could help aid in this problem, and thus improve release efficiency.

Future studies could explore the characteristics of release notes for different categories of open-source software, like Application Software and Framework. Studying a more advanced way to improve the Entry Summarizer and discriminate more change categories is also part of our future research agenda.

\section*{Acknowledgment}
This work was supported by the Strategy Priority Research Program of Chinese Academy of Sciences (No. XDA20080200).

\bibliographystyle{IEEEtran} 
\bibliography{mybib}

\end{document}